\documentclass[aps,prl,twocolumn,floatfix,showpacs,amsmath,amssymb]{revtex4}
\usepackage{graphicx}
\usepackage{dcolumn}
\usepackage{color} 
\begin{document}
\preprint{UCRL-JRNL-}
\title{$^7$Be(p,$\gamma$)$^8$B S-factor from \emph{ab initio} wave functions}
\author{P. Navr\'atil,$^1$ C.A. Bertulani,$^2$ and E. Caurier$^3$}
\affiliation{$^1$Lawrence Livermore National Laboratory, P.O. Box 808, L-414,
Livermore, CA  94551, USA \\
$^2$Department of Physics, University of Arizona, Tucson, AZ 85721, USA \\
$^3$Institut de Recherches Subatomiques (IN2P3-CNRS-Universit\'e Louis Pasteur)\\
            Batiment 27/1,67037 Strasbourg Cedex 2, France}

\date{\today}

\begin{abstract}
Nuclear structure of $^7$Be, $^8$B and $^{7,8}$Li is studied within
the {\it ab initio} no-core shell model (NCSM). Starting from the
high-precision CD-Bonn 2000 nucleon-nucleon (NN) interaction, wave functions of
$^7$Be and $^8$B bound states are obtained in basis spaces up to
$10\hbar\Omega$ and used to calculate channel cluster form factors
(overlap integrals) of the $^8$B ground state with $^7$Be+p.
Due to the
use of the harmonic oscillator (HO) basis, the overlap integrals
have incorrect asymptotic properties. 
We fix this problem in two alternative ways. First, by a
Woods-Saxon (WS) potential solution fit to the interior of the NCSM
overlap integrals.
Second, by a direct matching with the Whittaker function.
The corrected overlap integrals are then used for the
$^7$Be(p,$\gamma$)$^8$B S-factor calculation. We study the
convergence of the S-factor with respect to the NCSM HO frequency
and the model space size. Our
S-factor results are in agreement with recent direct measurement
data. 
\end{abstract}
\maketitle
%

%
The $^7$Be(p,$\gamma$)$^8$B capture reaction serves as an important input
for understanding the solar neutrino flux \cite{Adelberger}.
Recent experiments have determined the neutrino flux emitted from $^8$B
with a precision of ~9\% \cite{SNO}. On the other hand, theoretical predictions
have uncertainties of the order of 20\% \cite{CTK03,BP04}.
The theoretical neutrino flux depends on the $^7$Be(p,$\gamma$)$^8$B S-factor.
Many experimental and theoretical investigations studied this reaction.
Experiments were performed using direct techniques with proton beams
and $^7$Be targets \cite{Filippone,Baby,Seattle} as well as by indirect methods
when a $^8$B beam breaks up into $^7$Be and proton \cite{Be7pgamm_exp}.
Theoretical calculations needed to extrapolate the measured S-factor to the astrophysically
relevant Gamow energy were performed with several methods: the R-matrix parametrization
\cite{Barker95}, the potential model \cite{Robertson,Typel97,Davids03}, and the microscopic
cluster models \cite{DB94,Csoto95,D04}.

In this work, we present the first calculation of the
$^7$Be(p,$\gamma$)$^8$B S-factor starting from {\it ab initio}
wave functions of $^8$B and $^7$Be. We apply the {\it ab initio}
no-core shell model (NCSM) \cite{NCSMC12}. In this method, one considers nucleons
interacting by high-precision nucleon-nucleon (NN) potentials. There
are no adjustable or fitted parameters. 
We study
the binding energies and other nuclear structure properties of
$^7$Be, $^8$B as well as $^{7,8}$Li, and calculate overlap integrals
for the $^8$B and $^7$Be bound states. Due to the use of the
harmonic-oscillator (HO) basis, we have to correct the asymptotic
behavior of the NCSM overlap integrals. This is done in two alternative ways. 
First, by fitting
Woods-Saxon (WS) potential solutions to the interior part of the NCSM
overlap integrals under the constraint that the experimental
$^7$Be+p threshold energy is reproduced. 
Second, by a direct matching with 
the Whittaker function.
The corrected overlap integrals
are then utilized to calculate the $^7$Be(p,$\gamma$)$^8$B S-factor
as well as momentum distributions in stripping reactions. 

Our calculations for both $A=7$ and $A=8$ nuclei were performed 
using the high-precison CD-Bonn 2000 NN potential \cite{cdb2k} 
in model spaces up to $10\hbar\Omega$ ($N_{\rm max}=10$) for a wide range of HO frequencies. 
We then selected the optimal HO frequency
corresponding to the ground-state (g.s.) energy minimum in the $10\hbar\Omega$ space, 
here $\hbar\Omega=12$~MeV, and performed
a $12\hbar\Omega$ calculation to obtain the g.s. energy and the point-nucleon radii.
The overlap integrals as well as other observables were, however, calculated only using wave
functions from up to $10\hbar\Omega$ spaces.
The g.s. energies, radii and electromagnetic observables are summarized
in Table~\ref{tab:Be7moments}.
Note that the CD-Bonn 2000 underbinds $^7$Be, $^8$B and $^{7,8}$Li by about 3-5 MeV
and predicts $^8$B unbound, contrary to experiment. 
This suggests that the three-nucleon interaction is essential to accurately
reproduce the experimental threshold. 
However, since the HO basis has the incorrect asymptotic behavior in the first place, 
we make use of only the interior
part of our {\it ab initio} wave functions, which are likely unaffected by
mild variations in the threshold value. 
\begin{table}[hbtp]
  \caption{The $^7$Be and $^7$Li $\frac{3}{2}^- \frac{1}{2}$ ($^8$B and $^8$Li $2^+ 1$) 
g.s. energies (in MeV),
point-proton radii (in fm), quadrupole (in $e$fm$^2$)
and magnetic (in $\mu_{\rm N}$) moments and $\frac{1}{2}^-_1\rightarrow \frac{3}{2}^-_1$
($1^+_1\rightarrow 2^+_1$) M1 transitions (in $\mu_{\rm N}^2$)
obtained within the NCSM for the HO frequency of $\hbar\Omega=12$ MeV. 
Experimental values are from Refs. \cite{A=5-7,A=8}.
  \label{tab:Be7moments}}
  \begin{ruledtabular}
    \begin{tabular}{cccccc}
\multicolumn{3}{l}{$^7$Be} & \multicolumn{3}{l}{CD-Bonn 2000} \\
$N_{\rm max}$ & $|E_{\rm gs}|$ 
& $r_{\rm p}$ & Q & $\mu$ & B(M1)\\
\hline
 6 & 33.302 & 2.311 & -4.755 & -1.150 & 3.192 \\
 8 & 33.841 & 2.324 & -4.975 & -1.151 & 3.145 \\
10 & 33.972 & 2.342 & -5.153 & -1.141 & 3.114 \\
12 & 33.881 & 2.365 &  &  &\\
Expt. & 37.6004(5) & 2.36(2) &   & -1.398(15) & 3.71(48) \\
\hline
\multicolumn{3}{l}{$^7$Li} & \multicolumn{3}{c}{} \\
\hline
 6 & 34.951 & 2.149 & -2.717 & +3.027 & 4.256 \\
 8 & 35.494 & 2.156 & -2.866 & +3.020 & 4.188 \\
10 & 35.623 & 2.168 & -3.001 & +3.011 & 4.132 \\
12 & 35.524 & 2.188 & -3.130 & & \\
Expt. & 39.245 & 2.27(2) & -4.06(8) & +3.256 & 4.92(25) \\
\hline
\multicolumn{3}{l}{$^8$B} & \multicolumn{3}{c}{} \\
\hline
 6 & 31.772 & 2.436 & +5.218 & +1.463 & 3.498 \\
 8 & 32.258 & 2.463 & +5.420 & +1.455 & 3.506 \\
10 & 32.367 & 2.487 & +5.636 & +1.455 & 3.490 \\
12 & 32.284 & 2.520 & & & \\
Expt. & 37.7378(11) & 2.45(5) & (+)6.83(21) & 1.0355(3) & 9.1(4.5) \\
\hline
\multicolumn{3}{l}{$^8$Li} & \multicolumn{3}{c}{} \\
\hline
 6 & 35.352 & 2.139 & +2.588 & +1.238 & 4.454 \\
 8 & 35.834 & 2.139 & +2.690 & +1.243 & 4.428 \\
10 & 35.928 & 2.145 & +2.784 & +1.241 & 4.393 \\
12 & 35.820 & 2.161 & & & \\
Expt. & 41.277 & 2.26(2) & +3.27(6) & +1.654 & 5.01(1.61) \\
    \end{tabular}
  \end{ruledtabular}
\end{table}
Concerning the excitation energies, we obtained the same level ordering
for $^7$Be and $^7$Li.
Our CD-Bonn 2000 ordering is in agreement with experiment for the
9 lowest levels in $^7$Li. In $^7$Be, the experimental $7/2^-_2$ and
$3/2^-_2$ levels are reversed compared to our results and to the
situation in $^7$Li. 
While for the magnetic moments and M1 transitions we obtained a very small dependence 
of the calculated values on the HO frequency or the basis size, 
the radii and quadrupole moments
in general increase with increasing basis size and decreasing frequency.
The fastest convergence for the radii and quadrupole moment occurs at a smaller HO frequency. 
In our calculations with $\hbar\Omega=11$ and $12$ MeV, the radii are close 
to experimental values. 

From the obtained $^8$B and $^7$Be wave functions, we calculate the channel cluster 
form factors $g^{A\lambda J}_{(l\frac{1}{2})j;A-1 \alpha I_1}(r)$ following Ref.~\cite{cluster}.
Here, $A=8$, $l$ is the channel 
relative orbital angular momentum and 
$\vec{r}=\left[\frac{1}{A-1}
      \left(\vec{r}_1+\vec{r}_2 + \ldots+ \vec{r}_{A-1}\right)-\vec{r}_{A}\right]$
describes the relative distance between the proton and 
$^7$Be.
A conventional spectroscopic factor is obtained as
$S^{A\lambda J}_{(l\frac{1}{2})j;A-1 \alpha I_1}=
\int dr r^2
|g^{A\lambda J}_{(l\frac{1}{2})j;A-1 \alpha I_1}(r)|^2$.
Our selected spectroscopic factor results are summarized in
Table~\ref{tab:B8specfac}. We note a very weak dependence
of the spectroscopic factors on either the basis size or the HO
frequency.

\begin{table}[hbtp]
  \caption{The NCSM $\langle ^8$B$|^7$Be$+$p$\rangle$ spectroscopic factors.
  \label{tab:B8specfac}}
  \begin{ruledtabular}
    \begin{tabular}{ccccc}
\multicolumn{2}{c}{CD-Bonn 2000} & \multicolumn{3}{c}{$^7$Be+p $I_1^\pi (l,j)$} \\
$\hbar\Omega$ [MeV] & $N_{\rm max}$  &  $\frac{3}{2}^- (1,\frac{3}{2})$ &  $\frac{3}{2}^- (1,\frac{1}{2})$
&  $\frac{1}{2}^- (1,\frac{3}{2})$ \\
\hline
11 &  8 & 0.967 & 0.116 & 0.280 \\
11 & 10 & 0.959 & 0.111 & 0.275 \\
12 &  6 & 0.978 & 0.107 & 0.287 \\
12 &  8 & 0.969 & 0.103 & 0.281 \\
12 & 10 & 0.960 & 0.102 & 0.276 \\
14 &  8 & 0.967 & 0.085 & 0.284 \\
14 & 10 & 0.958 & 0.085 & 0.280 \\
    \end{tabular}
  \end{ruledtabular}
\end{table}

\begin{figure}[hbtp]
  \includegraphics*[width=0.9\columnwidth]
   {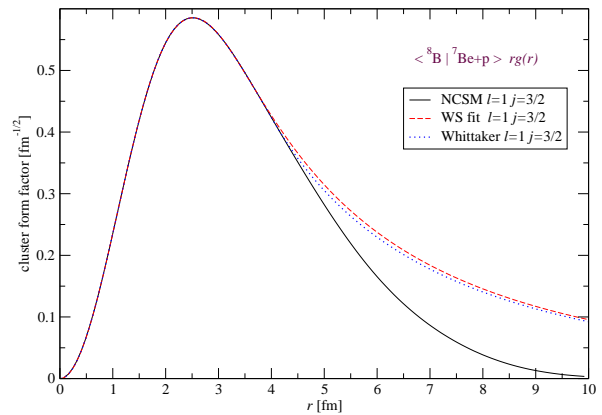}
  \caption{Overlap integral, $rg(r)$, for the g.s. of $^8$B with the g.s. 
of $^7$Be plus proton. For futher details see the text.
  \label{B8_Be7+p_overlap_12_3}}
\end{figure}

The dominant $p$-wave $j=3/2$ overlap integral for the $^8$B and $^7$Be ground states
obtained using the CD-Bonn 2000 NN potential in the $10\hbar\Omega$ model space
and the HO frequency of $\hbar\Omega=12$ MeV
is presented in 
Fig. \ref{B8_Be7+p_overlap_12_3} by the full line. Despite the fact
that a very large basis was employed, it
is apparent that the overlap integral is nearly zero at about 10 fm.
This is a consequence of the HO basis asymptotics. The proton
capture on $^7$Be to the weakly bound g.s. of $^8$B
associated dominantly by $E1$ radiation is a peripheral process.
Consequently, the overlap integral with an incorrect asymptotics
cannot be used to calculate the S-factor.
We expect, however, that the interior part of the overlap integral
is realistic. It is then straightforward
to correct its asymptotics. 
One possibility we explored utilizes solutions of a Woods-Saxon (WS) potential. 
In particular, we performed
a least-square fit of a WS potential solution to the interior of the
NCSM overlap in the range of $0-4$ fm. The WS potential parameters
were varied in the fit under the constraint that the experimental
separation energy of $^7$Be+p, $E_0=0.137$~MeV, was reproduced. In this way we obtain a perfect
fit to the interior of the overlap integral and a correct asymptotic behavior
at the same time. The result is shown in Fig. \ref{B8_Be7+p_overlap_12_3}
by the dashed line. 

Another possibility is a direct matching of logarithmic derivatives of the NCSM overlap integral
and the Whittaker function:
$\frac{d}{dr}ln(rg_{lj}(r))=\frac{d}{dr}ln(C_{lj} W_{-\eta,l+1/2}(2k_0r))$,
where $\eta$ is the Sommerfeld parameter, $k_0=\sqrt{2\mu E_0}/\hbar$ with $\mu$ the reduced mass
and $E_0$ the separation energy. 
Since asymptotic normalization constant (ANC) $C_{lj}$ cancels out, there is a unique solution at $r=R_m$.
For the discussed overlap presented in Fig.~\ref{B8_Be7+p_overlap_12_3}, we found $R_m=4.05$~fm.
The corrected overlap using the Whittaker function matching is shown in Fig.~\ref{B8_Be7+p_overlap_12_3}
by a dotted line. In general, we observe that the approach using the WS fit leads to deviations from the
original NCSM overlap starting at a smaller radius. In addition, the WS solution fit introduces
an intermediate range from about 4 fm to about 6 fm, where the corrected overlap deviates
from both the original NCSM overlap and the Whittaker function. Perhaps, this is a more realistic
approach compared to the direct Whittaker function matching. In any case, by considering the two alternative
procedures we are in a better position to estimate uncertainties in our S-factor results.

\begin{table}[hbtp]
  \caption{Parameters of the WS potentials obtained in the fits to the
interior part of the NCSM $\langle ^8$B$ (2^+_{\rm gs}) | ^7$Be$ (I_1^\pi) + $p$(l,j)\rangle (r)$
overlap functions.
  \label{tab:WSparam}}
  \begin{ruledtabular}
    \begin{tabular}{cccccccc}
\multicolumn{8}{c}{CD-Bonn 2000 $10\hbar\Omega$ $\hbar\Omega=12$ MeV} \\
$I_1^\pi (l,j)$ & $V_0$ & $R_0$ & $a_0$ & $V_{\rm ls}$ & $R_{\rm ls}$ & $a_{\rm ls}$ & $R_{\rm C}$ \\
\hline
$\frac{3}{2}^- (1,\frac{3}{2})$ & -51.037 & 2.198 & 0.602 & -9.719 & 2.964 & 0.279 & 2.198 \\
$\frac{3}{2}^- (1,\frac{1}{2})$ & -45.406 & 2.613 & 0.631 & -8.414 & 2.243 & 0.366 & 2.613 \\
$\frac{1}{2}^- (1,\frac{3}{2})$ & -49.814 & 2.235 & 0.553 & -17.024& 3.080 & 0.338 & 2.235 \\
\hline
\multicolumn{8}{c}{scattering state \protect\cite{Esbensen}} \\
\hline
  & -42.2 & 2.391 & 0.52 & -9.244 & 2.391 & 0.52 & 2.391 \\
    \end{tabular}
  \end{ruledtabular}
\end{table}
In the end, we re-scale the corrected overlap functions to preserve the original
NCSM spectroscopic factors (Table \ref{tab:B8specfac}).
The same procedure is applied to other relevant channels.
In Table~\ref{tab:WSparam}, we show examples of the fitted WS parameters
obtained in the two $p$-wave channels together with parameters corresponding to a $p$-wave
channel with $^7$Be in the first excited state ($E_0=0.57$ MeV).
We use the definition of the WS potential as given, e.g. in Eqs. (5-7) of Ref.~\cite{Radcap}.
Typically, the central potential parameters
$R_0$, $a_0$ are well constrained in the fit, while the spin-orbit potential parameters
are obtained with some uncertainty. 
The range used
in the least-square fit is not arbitrary but varies from channel to channel. 

The momentum distributions in break-up reactions present a direct
test of the corrected NCSM overlap integrals.
We compared our calculations performed as described in Ref.~\cite{BH04} 
to two sets of experimental data.
First, to an experiment performed at MSU for the
transverse momentum distributions of $^7{\rm Be}$ fragments from the
reaction $^8$B$+^9$Be at 41 MeV/A \cite{kel96}. 
Second, for the longitudinal momentum distributions, the data were taken
from a GSI experiment \cite{Gil02} which measured gamma rays in
coincidence with $^7{\rm Be}$ residues in the reaction $^8{\rm B}+^{12}$C
at 936 MeV/A. 
\begin{figure}[hbtp]
  \includegraphics*[width=0.8\columnwidth]
   {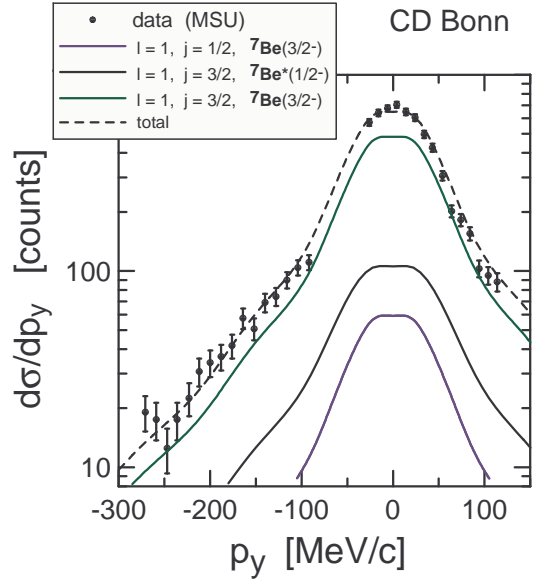}
\caption{Inclusive transverse-momentum distribution for the $^7$Be residue
in the $^{9}$Be($^{8}$B,$^{7}$Be)X reaction at 41 MeV/A \cite{kel96}. 
  \label{MSU_cdb2k}}
\end{figure}
Some of our results are summarized in Fig.~\ref{MSU_cdb2k} 
and Table~\ref{tab:B8cross2}. A good agreement with experiment is observed.
As the MSU data is given in arbitrary units, 
we multiplied the contributions 
from each channel by the same factor so that
their sum reproduces the maximum of the experimental distribution. 
\begin{table}[hbtp]
  \caption{Cross sections for the proton-removal reactions
$^{8}\mathrm{B}+\ ^{9}\mathrm{Be}$ at 41 MeV/nucleon (MSU) and
$^{8}\mathrm{B}+\ ^{12}\mathrm{C}$ at 936 MeV/nucleon (GSI). The
calculated total inclusive (excited state) cross sections, in mb, 
are given by $\sigma_{\rm inc}^{\left(\mathrm{th}\right)}$
($\sigma_{\rm exc}^{\left(\mathrm{th}\right)}$).
  \label{tab:B8cross2}}
\begin{ruledtabular}
\begin{tabular}{ccccc}
\multicolumn{5}{c}{CD-Bonn 2000 $10\hbar\Omega$ $\hbar\Omega=12$ MeV WS fit} \\
& $\sigma_{\rm inc}^{\left(\mathrm{exp}\right)}$ & $\sigma_{\rm inc}^{\left(\mathrm{th}\right)}$ 
& $\sigma_{\rm exc}^{\left(\mathrm{exp}\right)}$ & $\sigma_{\rm exc}^{\left(\mathrm{th}\right)}$ \\
\hline
MSU  & -- & 82.96 & -- & 15.31 \\
GSI  & $94\pm 9$ & 99.66 & $12\pm 3$ & 16.36 \\ 
\end{tabular}
\end{ruledtabular}
\end{table}

The S-factor for the reaction  $^7{\rm Be(p},\gamma)^8{\rm B}$
also depends on the continuum wave function,
$R_{lj}^{(c)}$. As we have not yet developed an
extension of the NCSM to describe continuum wave functions,
we obtain  $R_{lj}^{(c)}$ for $s$ and $d$ waves from
a WS potential model. 
Since the largest part of the integrand stays outside the
nuclear interior, one expects that the continuum wave functions are
well described in this way. 
In order to have the same scattering wave function in all the calculations,
we chose a WS potential from Ref.~\cite{Esbensen} that was fitted to
reproduce the $p$-wave $1^+$ resonance in $^8$B (Table~\ref{tab:WSparam}). 
It was argued \cite{Robertson}
that such a potential is also suitable for the description of $s$- and $d$-waves.
We note that the S-factor is very weakly dependent on the choice
of the scattering-state potential (using our fitted potential for the scattering state
instead changes the S-factor by less than 1.5 eV b at 1.6 MeV with no change at 0 MeV).

\begin{figure}[hbtp]
  \includegraphics*[width=0.9\columnwidth]
   {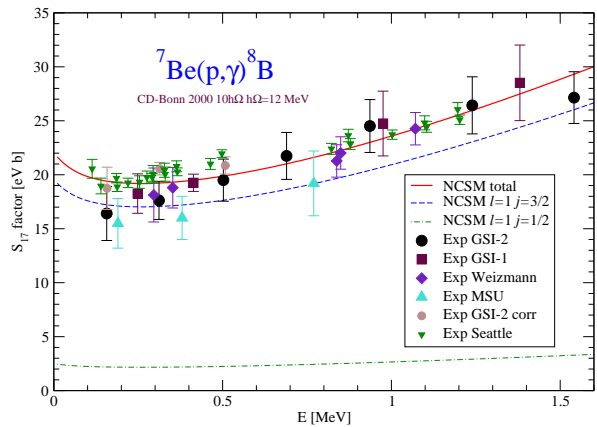}
  \caption{The $^7$Be(p,$\gamma$)$^8$B S-factor obtained
using the NCSM cluster form factors with corrected asymptotics
by the WS solution fit. 
Experimental values are from Refs. \protect\cite{Baby,Seattle,Be7pgamm_exp}.
  \label{S-factor_12_partial}}
\end{figure}

Fig. \ref{S-factor_12_partial} shows the astrophysical S-factor for the
reaction $^7{\rm Be(p},\gamma)^8{\rm B}$. We use bound-state wave functions
calculated with the CD-Bonn 2000 interaction in the $10\hbar\Omega$ model
space and the HO frequency $\hbar\Omega=12$ MeV.
The WS solution fit procedure was employed
to correct the asymptotics of the NCSM overlap functions.
The S-factor contributions from the dominant
$l=1$, $j=3/2$ and $j=1/2$ partial waves are drawn by the dashed lines. 
The full line is the sum of the two contributions.
The experimental data is a compilation of the latest experiments for the
S-factor. They include direct, as well as some indirect measurements
(Coulomb dissociation).
The slope of the curve corresponding to the total S-factor follows
the trend of the data.  
Our result is in a very good agreement with the recent direct measurement data
of Ref.~\cite{Seattle}.

\begin{table}[hbtp]
  \caption{The calculated $^7$Be(p,$\gamma$)$^8$B S-factor, in eV b, at the energy of 10 keV.
Two ways of correcting the NCSM overlap asymptotics, by the Woods-Saxon potential solution fit (WS)
and by a direct Whittaker function matching (Whit), are compared. The asymptotic normalization 
constants, in fm$^{-1/2}$, correspond to the Whittaker function matching case.    
  \label{tab:S-factor}}
  \begin{ruledtabular}
    \begin{tabular}{cccccc}
\multicolumn{6}{c}{CD-Bonn 2000} \\
$\hbar\Omega$ [MeV] & $N_{\rm max}$ & $S_{17}^{\rm WS}(10)$ & $S_{17}^{\rm Whit}(10)$ 
& $C_{1, 3/2}$ & $C_{1, 1/2}$ \\
\hline
15 & 6  & 17.80 & 16.81 & 0.647 & 0.195   \\
15 & 8  & 18.87 & 17.58 & 0.660 & 0.206   \\
15 & 10 & 19.81 & 18.33 & 0.672 & 0.216   \\
14 & 10 & 20.21 & 18.78 & 0.680 & 0.220   \\
13 & 10 & 21.02 & 19.64 & 0.692 & 0.234   \\
12 & 6  & 21.24 & 19.75 & 0.693 & 0.240   \\
12 & 8  & 21.14 & 19.96 & 0.696 & 0.242   \\
12 & 10 & 21.66 & 20.45 & 0.704 & 0.247   \\
11 & 6  & 22.38 & 21.30 & 0.715 & 0.261   \\
11 & 8  & 23.04 & 21.33 & 0.715 & 0.263   \\
11 & 10 & 23.06 & 21.60 & 0.720 & 0.262   \\
\multicolumn{2}{c}{NCSM $S_{17}(10\;{\rm keV})$} & \multicolumn{2}{c}{$22.1\pm 1.0$} & & \\
    \end{tabular}
  \end{ruledtabular}
\end{table}

In order to judge the convergence of our S-factor calculation, we performed
a detailed investigation of the model-space-size and the HO-frequency dependencies.
In Table~\ref{tab:S-factor}, we summarize our
S-factor results at 10 keV. We observe a steady increase of the S-factor with
the basis size enlargement for higher frequencies. Contrary to this situation, the
calculation using the HO frequency of $\hbar\Omega=11$ MeV and the WS solution fit
shows that the S-factor does not increase any more with increasing $N_{\rm max}$.
We also present the $S_{17}$ and the ANC obtained using the alternative
direct Whittaker matching procedure. In general, both procedures lead to basically
identical energy dependence with a difference of about 1 to 2 eV b in the S-factor
with the smaller values from the direct Whittaker function matching procedure.
Results at $\hbar\Omega=11$ and 12 MeV show very weak dependence on $N_{\rm max}$, 
with relative difference between the two methods always in the range of 5 to 8\%.
The full range of results is covered by 
$S_{17}(10\;{\rm keV})=22.1\pm 1.0$ eV b.
We stress that no adjustable parameters were used in our
{\it ab initio} calculations of the $^8$B and $^7$Be bound states. Taking into account
that the S-factor is only weakly dependent on the potential model used to obtain
the scattering state, we consider our results as the first {\it ab initio} prediction
of the $^7$Be(p,$\gamma$)$^8$B S-factor, in particular of its normalization.

We thank W. E. Ormand, C. Forss\'en and J. Dobe\v{s} for useful discussions.
This work was partly performed under the auspices of the
U. S. Department of Energy by the University of California, Lawrence
Livermore National Laboratory under contract No. W-7405-Eng-48. Support
from the LDRD contract No.~04--ERD--058, from 
U.S. DOE, OS (Work Proposal Number SCW0498),
and  grant No. DE-FG02-04ER41338, is acknowledged.

\end{document}